\documentclass[aps,prd,twocolumn,superscriptaddress,floatfix,amsmath,longbibliography,nofootinbib,amssymb,raggedbottom]{revtex4-2}
\usepackage[colorlinks=true,citecolor=blue,linkcolor=blue]{hyperref}
\usepackage{graphicx}
\usepackage{needspace}
\usepackage{subfigure}
\subfiguretopcaptrue
\subcapraggedrighttrue
\subcapnoonelinetrue
\newlength{\panelcaptionleft}
\makeatletter
\renewcommand{\@makesubfigurecaption}[2]{%
  \hbox to\@tempdima{%
    \hspace*{\panelcaptionleft}%
    \parbox[t]{\dimexpr\@tempdima-\panelcaptionleft\relax}{%
      \subcapsize\raggedright
      {\subcaplabelfont #1}{\subcapfont\ignorespaces #2}}\hss}}
\makeatother
\usepackage{dcolumn}
\usepackage{mathrsfs}
\usepackage{bm}
\usepackage{amssymb}
\hypersetup{pdfauthor={Yu Shi, Sheng Long},hypertexnames=false}
\begin{document}
\raggedbottom
\title{Boundary preparation and anomalous relaxation near a black hole exceptional point}
\author{Yu Shi}
\email{shiyu@hhtc.edu.cn}
\affiliation{Department of Physics, Huaihua University, Huaihua, Hunan 418008, China}

\author{Sheng Long}
\email[Corresponding author: ]{shenglong@ucas.ac.cn}
\affiliation{School of Fundamental Physics and Mathematical Sciences, Hangzhou Institute for Advanced Study, University of Chinese Academy of Sciences, Hangzhou 310024, China}
\begin{abstract}
We study the relaxation of a Robin scalar field near a stable exceptional point of a planar Schwarzschild AdS black hole, where finite boundary driving controls the recovery time without changing the leading decay rate at linear order. At the exceptional point, two weighted moments of the source determine the time dependence of the leading response. When the zeroth moment vanishes while the first remains nonzero, the linear time factor is eliminated, leaving an ordinary exponential decay. With suitable boundary preparation, a perturbation that retains more energy at release can nevertheless reach a common recovery threshold sooner. At the exceptional point, its recovery time advantage over a fixed reference that retains the linear factor grows as \(\gamma_*^{-1}\log\log(1/\theta)\) as the common amplitude threshold \(\theta\) approaches zero, where \(\gamma_*\) is the common decay rate. The source moments also determine the constant term in this asymptotic time difference. Full scalar field evolution on a fixed background near the exceptional point illustrates the reversal for pulses matched in duration, boundary task, source cost, and injected work, within the time window where the small spectral splitting remains unresolved. The injected work also modifies the final spectrum through gravitational heating. When weak driving begins with a black hole at the exceptional point, the temperature rise is quadratic in the source amplitude, whereas the final pole splitting is linear in its magnitude. This splitting limits the time window over which the exceptional point recovery law can be observed.

\end{abstract}
\maketitle

\section{Introduction}
Under certain conditions, hot water can freeze faster than cold water, a counterintuitive phenomenon known as the Mpemba effect \cite{MpembaOsborne1969}. More generally, a system initially farther from equilibrium can reach a given recovery threshold sooner than a closer system governed by the same relaxation dynamics \cite{C01,C02,C04,C05,TezaEtAl2026Review}. Controlled experiments with colloids and trapped ions have demonstrated this reversal in the order of recovery \cite{C03,C06}. To understand it, one must look beyond the initial distance from equilibrium and examine how the perturbations decay. Near equilibrium, relaxation can often be decomposed into modes with different decay rates, and the initial state determines how strongly each mode is excited. A state farther from equilibrium may predominantly excite rapidly decaying modes, while a closer state can retain a larger contribution from slow modes. Preparation can therefore offer a way to control the order of recovery by changing the weights of these modes \cite{MoroderEtAl2024Thermodynamics,HongEtAl2026Steering}.

As a parameter varies in an open system, eigenvalues and their corresponding eigenvectors can coalesce at an exceptional point (EP). Such a merger affects relaxation through both the decay rates and the time dependence of the response. Suitable Liouvillian EPs have been shown to increase the spectral decay rate and accelerate relaxation \cite{ZhouEtAl2023LiouvillianEP}. At an EP of order two, a factor linear in time can multiply the exponential decay, and the resulting Jordan time dependence can produce Mpemba crossings \cite{MP01}. Other work has shown that exchanging reservoirs can select a pure exponential response or one that retains the time factor while leaving the spectrum unchanged \cite{YanEtAl2026Nonreciprocal}. Even when two perturbations share the same exponential decay rate, a linear time factor can delay recovery. Can a pulse of finite duration control this delay when both perturbations evolve under the same dynamics after release?

Black holes are among the most remarkable systems in nature, and their macroscopic simplicity makes them a natural laboratory for studying this question. The exterior geometry of a typical stationary black hole can be described by a few macroscopic parameters. Its horizon determines its temperature and entropy and provides a clear channel through which exterior perturbations are absorbed. The energy supplied during preparation can also alter the black hole geometry through gravitational backreaction, changing the background on which perturbations subsequently relax \cite{SbernaEtAl2022Absorption}. On this thermodynamic background, quasinormal modes can be used to study the decay of perturbations and connect their evolution to energy absorption. Recent studies of black hole EPs have progressed from identifying mode coalescence to explaining resonant excitation near coalescence and its effects on the subsequent ringdown \cite{CavalcanteEtAl2024KerrEP,G14,OshitaEtAl2025Destructive,YangEtAl2025Resonances,G09,NakamotoOshita2026Nariai}. Interference between nearly coalescing modes can change the ringdown waveform, while the manifestation of an EP in a physical response also depends on how the perturbation is excited and which observable is measured \cite{MorikawaEtAl2026RieszLaurent,Destounis2026RingdownEP}. Accounting for the preparation process is therefore a natural step toward understanding relaxation near a black hole EP.

Recent studies have examined black hole EPs with Robin boundary conditions \cite{WangYang2026Robin} and their responses to finite boundary driving \cite{LongEtAl2026BTZ}. We consider a Robin scalar in the background of a planar Schwarzschild AdS black hole, where the leading modes coalesce within the stable phase at a finite decay rate \cite{R1}. This setting allows us to examine the effect of the EP time structure on recovery in a stable system with sustained dissipation. Finite pulses applied at the AdS boundary can prepare perturbations \cite{HP01,R6}. Once the pulses are removed, the perturbations evolve under the same equations and boundary conditions, so their recovery can be compared directly. Through holography \cite{G01,G02,G03}, this process can also be understood as the return to equilibrium of a strongly coupled thermal state after an external drive is removed \cite{G04,G05,R2}. Related holographic studies have used horizon flux to track recovery in a superfluid \cite{GeEtAl2026HolographicMpemba} and examined how boundary drives prepare states that relax anomalously in a superconductor \cite{ShiLong2026EntropyCharge}.

To study how preparation history changes the order of recovery, we connect mode selection by finite boundary pulses to the balance of energy. Some of the energy injected by a pulse is absorbed by the horizon during preparation, while the rest remains outside the black hole. We measure recovery using the total exterior scalar energy, including the Robin boundary term. In the stable regime considered here, this energy is positive definite and decreases monotonically after release. Our analytic solution shows that a suitable pulse can remove the linear time factor from the leading response while preserving the leading exponential decay rate. With an appropriate preparation, a perturbation that retains more energy at release can then reach a common recovery threshold sooner than a reference perturbation that retains the time factor. At the exact EP, the time advantage for a fixed pair of preparations grows asymptotically as \(\gamma_*^{-1}\log\log(1/\theta)\) as the common amplitude threshold \(\theta\) decreases, where \(\gamma_*\) is the common decay rate.

Full scalar field evolution near the EP illustrates the reversal for pulses with matched preparation resources and equal injected work. An analysis of the leading gravitational backreaction also shows that heating by the drive changes the final temperature and quasinormal mode spectrum of the black hole. The resulting spectral splitting limits the observation window for the EP recovery law. Thus, both the selection of modes and the change in the background caused by preparation affect the observability of the recovery advantage.

The remainder of this paper is organized as follows. Section~\ref{sec:preparation} formulates the boundary preparation protocol and the associated energy balance. Section~\ref{sec:source-memory} analyzes how weighted source moments select the time dependence of the coalesced mode response and affect the fractional rate of horizon absorption. Section~\ref{sec:recovery} derives the resulting recovery time advantage and illustrates the energy reversal numerically near the exceptional point. Section~\ref{sec:detuning} examines detuning, gravitational heating, and dynamical entropy production, and Sec.~\ref{sec:discussion} concludes with a summary and outlook. Appendix~\ref{app:radial-preparation} presents the radial analysis and the details of pulse construction and field evolution.

\section{Preparation protocol and energy balance}
\label{sec:preparation}
We begin with a scalar treated as a linear probe on a fixed black hole. Starting from zero, the field is driven by a real source supported on \([-\tau_p,0]\) that vanishes at both endpoints. The source is removed at \(t=0\), after which the field relaxes freely. We compare preparations with the same duration on the same background, taking the planar Schwarzschild AdS$_4$ black hole geometry
\begin{equation}
\begin{gathered}
ds^2=\frac{L^2}{z^2}\left[-Fdt_{\rm Sch}^2+\frac{dz^2}{F}
+dx_1^2+dx_2^2\right],\\
F=1-(z/z_h)^3,\qquad T=\frac3{4\pi z_h}.
\end{gathered}
\label{eq:metric}
\end{equation}
We use a real homogeneous scalar of mass \(\mu^2L^2=-2\). Its Schwarzschild boundary expansion is \(\Phi=z\alpha+z^2\beta+\cdots\), and the physical Robin source is \(J=\beta-\kappa_{\rm phys}\alpha\). We fix \(\kappa_{\rm phys}<0\), with \(q=|\kappa_{\rm phys}|\). The Robin variational problem has boundary term \(-\mathcal N\int(\kappa_{\rm phys}\alpha^2/2+J\alpha)\) and expectation value \(-\mathcal N\alpha\), where \(\mathcal N>0\) \cite{G06,G07,G08,G10,G11}. In this quantization, the quadratic term in \(\alpha\) implements a double trace deformation of the boundary theory. The coupling \(\kappa_{\rm phys}\) is held fixed in both preparations. Only the external source \(J\) varies with time, and it vanishes after release. The renormalized Ward identity then gives the work from the variation of \(J\) and the response conjugate to it \cite{BianchiFreedmanSkenderis2002Renorm,IshiiEtAl2022Extraction}.

Because energy leaves the exterior through the horizon, it is convenient to use \(x=\int_0^z d\zeta/F(\zeta)\) and the ingoing Eddington--Finkelstein time \(t=t_{\rm Sch}-x\). This time agrees with boundary time and is regular at the horizon, while its constant surfaces are ingoing null slices on which the horizon flux can be evaluated. Writing \(\Phi=za(t)+z^2b(t)+\cdots\) in these coordinates gives \(\alpha=a\) and \(\beta=b-\dot a\). With \(f=\Phi/z\) and the positive potential \(V(x)=F(z(x))z(x)/z_h^3\), the field equation and the driven Robin condition are
\begin{equation}
\begin{aligned}
f_{xx}-2f_{tx}-Vf &=0,\\
f_x(t,0)-f_t(t,0)-\kappa_{\rm phys}f(t,0)&=J(t).
\end{aligned}
\label{eq:field-source}
\end{equation}
The boundary time derivative is the time domain form of the frequency term in the Eddington--Finkelstein (EF) Robin condition of Ref.~\cite{R1}. We denote the regular horizon value by \(f_H\).

On an ingoing null slice the exterior scalar energy includes the full radial profile and the boundary term fixed by the Robin action \cite{R1}.

\begin{equation}
2\mathcal E[f]=\int_0^\infty\left(|f_x|^2+V|f|^2\right)dx
+\kappa_{\rm phys}|f(0)|^2.
\label{eq:energy}
\end{equation}
Because \(\kappa_{\rm phys}<0\), the boundary term in \eqref{eq:energy} is negative and the positivity of the energy is not manifest. For Robin problems one can make positivity explicit by weighting the radial derivative with a suitable static profile \cite{HolzegelWarnick2014Boundedness}. We use the positive regular static solution of Ref.~\cite{R1}, normalized to one at the horizon, which obeys \(u_c''=Vu_c\) and \(u_c'(0)=\kappa_cu_c(0)\). Its critical coupling is
\(\kappa_c=\bar\kappa_c/z_h\), with
\(\bar\kappa_c=-3\Gamma(2/3)^3/\Gamma(1/3)^3\) \cite{RenXie2023ZeroDensity}.
An integration by parts gives
\begin{equation}
2\mathcal E[f]=\int_0^\infty u_c^2
\left|\partial_x(f/u_c)\right|^2dx
+(\kappa_{\rm phys}-\kappa_c)|f(0)|^2.
\label{eq:positive-energy}
\end{equation}
The horizon boundary term vanishes because \(u_c'\) decays exponentially. The characteristic energy is therefore positive on the stable side \(\kappa_{\rm phys}>\kappa_c\), which includes the EP studied below.

Horizon flux also governs the energy of a Robin scalar in spherical Schwarzschild AdS \cite{FicekMaliborski2024Robin}. For the driven planar problem, multiplying the bulk equation in \eqref{eq:field-source} by \(f_t\) and integrating over \(x\) gives
\begin{equation}
\dot{\mathcal E}=-J\dot\alpha-|\dot f_H|^2.
\label{eq:driven-balance}
\end{equation}
In this calculation, the mixed derivative \(2f_{tx}\) gives the endpoint values of \(f_t^2\). The contribution at \(x=0\) combines with the radial derivative term by the driven Robin condition, leaving the power \(-J\dot\alpha\) after the Robin boundary energy is included. At the horizon, regularity leaves the absorbed flux \(|\dot f_H|^2\). Integrating during preparation gives the injected work per transverse area, \(W_{\rm inj}/\mathcal A=\mathcal N\mathcal W\), where
\begin{equation}
\begin{split}
\mathcal W[J]&=\int\dot J\alpha\,dt=-\int J\dot\alpha\,dt\\
&=\mathcal E[f_J(0)]+\int_{-\tau_p}^0|\dot f_{H,J}|^2dt.
\end{split}
\label{eq:work-partition}
\end{equation}
The two work integrals agree because the source vanishes at the endpoints. Equal work can leave unequal release energies because the horizon can absorb different amounts during the two pulses.

After release the horizon absorbs the remaining energy. For a field that decays completely, integrating \eqref{eq:driven-balance} to the final state gives
\begin{equation}
\mathcal E[f_J(t)]=\int_t^\infty|\dot f_{H,J}(u)|^2du,\qquad t\ge0.
\label{eq:remaining-absorption}
\end{equation}
Once the source turns off, no further work is done on the system, and \(\mathcal E[f_J(t)]\) measures the remaining exterior energy to be absorbed by the horizon. As shown in Sec.~\ref{sec:detuning}, at the first nonzero gravitational order this energy is proportional to the remaining local dynamical entropy deficit of the black hole. Equation~\eqref{eq:entropy-deficit} gives the normalization in terms of the temperature and the scalar coupling.

A comparison of preparations must account for both the target achieved and the resources used to reach it. We therefore prescribe the same average boundary response and match both the source cost and the work in \eqref{eq:work-partition}. The task and cost are
\begin{equation}
\begin{gathered}
Q[J]=\frac1\Delta\int_0^\Delta\alpha_J(t)dt=A_0\ne0,\\
\mathcal B[J]=\int_{-\tau_p}^0\left(J^2+\ell_b^2\dot J^2\right)dt.
\end{gathered}
\label{eq:comparison}
\end{equation}
The positive scales \(\tau_p,\Delta,\ell_b\) are fixed. The cost weights the amplitude and rate of change of the drive. The waveform and its conjugate response determine the work independently of this cost.

We compare the exterior fields in the common energy unit \(E_u=K_{\rm stat}A_0^2/2\), using the dimensionless amplitude
\begin{equation}
D_J(t)=\sqrt{\frac{2\mathcal E[f_J(t)]}{K_{
\rm stat}A_0^2}},\qquad
K_{\rm stat}=\kappa_{\rm phys}-\kappa_c>0.
\label{eq:recovery-amplitude}
\end{equation}
The average in \eqref{eq:comparison} guarantees that \(|\alpha_J(t_*)|\ge |A_0|\) at some \(t_*\in[0,\Delta]\). Equation~\eqref{eq:positive-energy} bounds the energy on that slice from below by \(K_{\rm stat}|\alpha_J(t_*)|^2/2\). Because the energy decreases after release, \(\mathcal E[f_J(0)]\ge \mathcal E[f_J(t_*)]\ge E_u\). Thus the response required by the task supplies a common lower bound on the release energy and a common scale for comparing the two pulses.

The recovery time \(\tau_J\) is the first time at which \(D_J\le\theta\). Since the energy decreases after release, it stays below the threshold thereafter. Including preparation gives the total time \(\tau_p+\tau_J\), as in operational comparisons of Mpemba acceleration \cite{NavaEgger2025Pontus}. We vary \(\theta\) with both waveforms fixed.

While the partition of work during preparation sets the energy left at release, the subsequent recovery depends on which modes the source history excites.

\section{Source memory at the exceptional point}
\label{sec:source-memory}
\begin{figure*}[t]
\centering
\begin{minipage}[t]{.610\textwidth}\vspace{0pt}
{\setlength{\panelcaptionleft}{0.116152\linewidth}
\setlength{\subfigcapskip}{\dimexpr 3pt-0.019359\linewidth\relax}
\subfigure[Energy ordering]{\includegraphics[width=\linewidth]{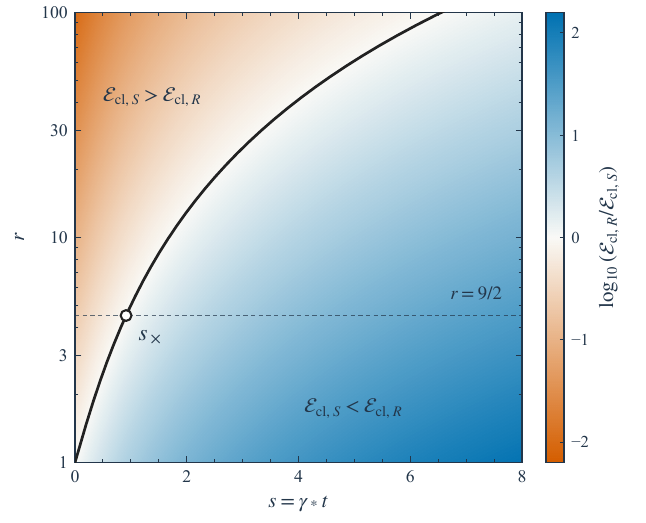}\label{fig:F1a}}}
\end{minipage}
\hfill
\begin{minipage}[t]{.370\textwidth}\vspace{0pt}
{\setlength{\panelcaptionleft}{0.225963\linewidth}
\setlength{\subfigcapskip}{\dimexpr 3pt-0.031916\linewidth\relax}
\subfigure[Horizon memory]{\includegraphics[width=\linewidth]{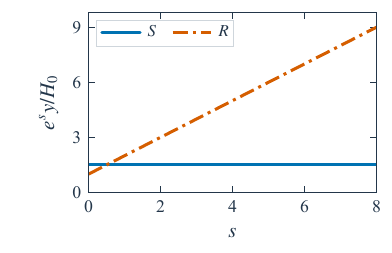}\label{fig:F1b}}}
\par\vspace{1.5pt}
{\setlength{\panelcaptionleft}{0.225963\linewidth}
\setlength{\subfigcapskip}{\dimexpr 3pt-0.031916\linewidth\relax}
\subfigure[Fractional absorption]{\includegraphics[width=\linewidth]{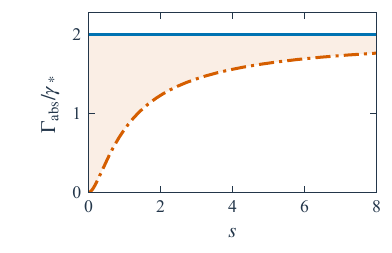}\label{fig:F1c}}}
\end{minipage}
\caption{Energy reversal in the isolated leading EP cluster. Panel~\subref{fig:F1a} shows \(\log_{10}(\mathcal E_{{\rm cl},R}/\mathcal E_{{\rm cl},S})\) for \eqref{eq:cluster-reversal}, with initial energy ratio \(r>1\). Warm colors show the initial energy ordering and cool colors show its reversal. The black curve marks equal energies. The dashed path selects \(r=9/2\), with the crossing marked by an open circle. Panels~\subref{fig:F1b} and~\subref{fig:F1c} show this pair. Removing the common exponential exposes the linear horizon memory of \(R\). This memory reduces its fractional absorption rate by the shaded amount. The legend in~\subref{fig:F1b} applies to both curves in~\subref{fig:F1c}.}
\label{fig:mechanism}
\end{figure*}

\subsection{Finite decay at the mode merger}

We use the Laplace convention in which decaying quasinormal modes have \(\operatorname{Re}p<0\). Let \(G_p\) be the regular solution of \(G_p''-2pG_p'-VG_p=0\), normalized by \(G_p(\infty)=1\). The boundary response is
\begin{equation}
\begin{gathered}
\mathcal R(p)=\frac{G_p(0)}{H_\kappa(p)}
=-\frac1{m(p)+\kappa_{\rm phys}},\\
H_\kappa(p)=G_p'(0)-(p+\kappa_{\rm phys})G_p(0),\\
m(p)=p-\frac{G_p'(0)}{G_p(0)}.
\end{gathered}
\label{eq:response}
\end{equation}
The merger at \(\kappa_*=-m(-\gamma_*)\) is a leading double pole with a finite decay rate \(\gamma_*>0\). The coupling satisfies \(\kappa_c<\kappa_*<0\). At fixed negative physical coupling the merger temperature \(T_*\) is therefore above the zero mode threshold \(T_c=3q/(4\pi|\bar\kappa_c|)\). The radial argument in Appendix~\ref{app:radial-preparation} gives \(m'(-\gamma_*)=0\), \(m''(-\gamma_*)>0\) and \(\gamma_*<\nu/2\), where \(\nu=3/z_h\) is the decay rate of the horizon tail of the radial potential. The merger thus occurs inside the stable phase.

At coalescence the response takes the form
\begin{equation}
\begin{gathered}
\mathcal R(p)=\frac{\mathsf b_*}{(p+\gamma_*)^2}
+\frac{\mathsf a_*}{p+\gamma_*}+\cdots,\\
\mathsf b_*=-\frac2{m''(-\gamma_*)}<0.
\end{gathered}
\label{eq:double-pole}
\end{equation}
A double pole produces a linear time factor \cite{E03,R3}, whose coefficient depends on excitation and readout \cite{Wiersig2022ResponseStrengths,MorikawaEtAl2026RieszLaurent,LiuWu2026PlungeEP}. Constrained initial perturbations have been used to suppress this term in a curvature signal at fixed Kerr--Newman--de Sitter geometry \cite{Destounis2026RingdownEP}. Here two weighted time integrals of the finite Robin pulse determine the leading response after release.

\subsection{Two moments of the source history}
Convolving the pole contribution \((\mathsf a_*+\mathsf b_*t)e^{-\gamma_*t}\) with the source shifts the time argument of the kernel to \(t-s\), where \(s\) is the source time. Separating the dependence on \(s\) gives two weighted moments,
\begin{equation}
\begin{gathered}
M_0[J]=\int_{-\tau_p}^0e^{\gamma_*s}J(s)ds,\\
M_1[J]=\int_{-\tau_p}^0s e^{\gamma_*s}J(s)ds.
\end{gathered}
\label{eq:source-moments}
\end{equation}
The contribution of the merged poles to the released boundary response is
\begin{equation}
\begin{gathered}
\alpha_{{\rm cl},J}(t)=(C_J+B_Jt)e^{-\gamma_*t},\\
B_J=\mathsf b_*M_0[J],\qquad
C_J=\mathsf a_*M_0[J]-\mathsf b_*M_1[J].
\end{gathered}
\label{eq:source-memory}
\end{equation}

The weight \(e^{\gamma_*s}\) accounts for the exponential decay of each excitation between its application and release. Since the weight is positive, setting \(M_0=0\) with a nonzero real pulse requires cancellation between its positive and negative parts. The time weighted moment \(M_1\) can remain nonzero. This choice selects the time dependence without moving the poles of the system. The underlying idea of placing a zero in the source transform is familiar from input shaping \cite{R4,R5,C07,C08}. The transform \(\widehat J(p)=\int_{-\tau_p}^0e^{-ps}J(s)ds\) obeys
\[
\widehat J(-\gamma_*)=M_0[J],\qquad
\widehat J'(-\gamma_*)=-M_1[J].
\]
When \(M_0=0\) and \(M_1\ne0\), the simple zero of the source reduces the pole of the product \(\mathcal R\widehat J\) to order one at \(-\gamma_*\). The system response \(\mathcal R\) retains its double pole. The prepared response loses its time factor and keeps the same exponential rate. Robust vibration control uses additional zeros to suppress residual excitation under detuning \cite{SinghVadali1995Robust}. Here imposing a double zero would also set \(M_1=0\) and cancel the ordinary exponential.

The source moment can also be identified from the resulting time signal, since polynomial coefficients at an EP can be extracted from the response \cite{FuchsEtAl2014HarmonicInversion}. At exact coalescence the compensated boundary signal \(e^{\gamma_*t}\alpha_J(t)\) approaches a straight line with slope \(\mathsf b_*M_0[J]\). Varying the pulse through \(M_0=0\) changes the sign of the slope. At zero slope a nonzero \(M_1\) leaves a finite intercept.

\subsection{Memory and fractional absorption}
Equation~\eqref{eq:remaining-absorption} relates the remaining exterior energy to future horizon flux, so we now examine how the source history appears in the leading horizon field. Write \(\gamma=\gamma_*\) and let \(y=e^{-\gamma t}(A_H+B_Ht)\) denote the leading horizon field. Its coefficients determine those of the boundary response through
\[
B=gB_H,\qquad C=gA_H+hB_H,
\]
where \(g=G_{-\gamma}(0)>0\) and \(h=\partial_pG_p(0)|_{-\gamma}\).
Thus cancelling \(M_0\) removes the time factor at the horizon as well.

With \(G_*=G_{-\gamma}\), the associated radial field is
\[
f_{\rm cl}=e^{-\gamma t}\left[(A_H+B_Ht)G_*+
B_H\left.\partial_pG_p\right|_{-\gamma}\right].
\]
The term proportional to \(\partial_p G_p\) comes from differentiating the radial solution together with its time factor. Applying the Robin operator to \(e^{pt}G_p\) gives \(e^{pt}H_\kappa(p)\), so both the mode and its frequency derivative satisfy the homogeneous condition at the double zero. The isolated cluster therefore obeys the exact balance \(\dot{\mathcal E}_{\rm cl}=-|\dot y|^2\). Since it decays completely, its exterior energy equals the horizon flux still to be absorbed.
\begin{equation}
\mathcal E_{\rm cl}(t)=\int_t^\infty|\dot y(u)|^2du
=\frac{\dot y(t)^2+\gamma^2y(t)^2}{4\gamma}.
\label{eq:cluster-energy}
\end{equation}
To obtain the last expression, use \(\ddot y+2\gamma\dot y+\gamma^2y=0\). The derivative of the quadratic form is then \(-\dot y^2\), and the form vanishes as \(t\to\infty\). Thus the energy of the leading radial response has the same balance as a critically damped oscillator and can be computed from the horizon field alone. Finite energy thresholds also give an operational measure of oscillator recovery \cite{LelasPoljakJukic2023Oscillator}. Here the pulse sets \(y\) and \(\dot y\) at a common damping rate.

Let \(S\) denote a nonzero pure exponential and \(R\) a reference with a nonzero Jordan coefficient. Their fractional absorption rates \(\Gamma_{\rm abs}=-\partial_t\log\mathcal E_{\rm cl}\) obey
\begin{equation}
\begin{aligned}
\Gamma_{{\rm abs},S}&=2\gamma,\\
\Gamma_{{\rm abs},R}&=2\gamma-\frac{2}{t}+O(t^{-2}).
\end{aligned}
\label{eq:absorption-rate}
\end{equation}
The time factor slows the fractional loss of the remaining energy to the horizon, so \(S\) has the larger fractional absorption rate at late times. This excess decreases as \(2/t\), but its time integral grows logarithmically. Since this integral is the change in \(\log(\mathcal E_{{\rm cl},R}/\mathcal E_{{\rm cl},S})\), the energy ratio grows quadratically in time. The accumulated absorption difference can therefore reverse the initial ordering while both responses keep the same exponential rate.

To see how this difference reverses the energy ordering, consider the isolated cluster family
\(y_R=H_0(1+s)e^{-s}\) and \(y_S=aH_0e^{-s}\), where
\(s=\gamma_*t\), \(H_0\ne0\) and \(a>1/\sqrt{2}\).
Its initial energy ratio is
\(r=\mathcal E_{{\rm cl},S}(0)/\mathcal E_{{\rm cl},R}(0)=2a^2>1\).
Equation~\eqref{eq:cluster-energy} gives
\begin{equation}
\begin{gathered}
\frac{\mathcal E_{{\rm cl},R}(t)}{\mathcal E_{{\rm cl},S}(t)}
=\frac{1+2s+2s^2}{r},\\
s_\times=\frac{\sqrt{2r-1}-1}{2}.
\end{gathered}
\label{eq:cluster-reversal}
\end{equation}
The energy ratio grows from \(1/r<1\), crosses unity at \(s_\times\) and stays above it thereafter. The common exponential cancels from the ratio. The reversal in Fig.~\ref{fig:mechanism} is set by the polynomial memory of the reference response. Section~\ref{sec:fixed-example} demonstrates this mechanism near coalescence with pulses satisfying the common preparation conditions.

\section{Energy reversal and recovery times}
\label{sec:recovery}
\subsection{Recovery times from source memory}
\label{sec:recovery-times}
We now examine when source memory reverses the order of recovery for a fixed pair of preparations at exact coalescence that satisfy the common conditions of Sec.~\ref{sec:preparation}. Their release energies obey \(\mathcal E_{0,S}>\mathcal E_{0,R}\). Source \(S\) has \(M_0[J_S]=0\) and \(M_1[J_S]\ne0\), while \(M_0[J_R]\ne0\). We compare their recovery using the common energy amplitude \(D_J\).

The quasinormal spectrum alone leaves the size and duration of transients undetermined \cite{Warnick2015AdSQNM,CarballoWithers2024Transients}. We establish dominance of the leading cluster by bounding the remaining radial field. For a fixed source of finite duration and cost,
\begin{equation}
\begin{gathered}
f_J=f_{{\rm cl},J}+r_J,\qquad
\|r_J(t)\|_{\mathcal E}\le C\|J\|_{L^2}e^{-a_rt},\\
\|f\|_{\mathcal E}=\sqrt{2\mathcal E[f]},\qquad
\gamma_*<a_r<\nu/2.
\end{gathered}
\label{eq:full-remainder}
\end{equation}
After extracting the double pole, we shift the inverse Laplace contour to \(\operatorname{Re}p=-a_r\) within the spectral gap above the next pole. On this contour the radial transfer \(G_p/H_\kappa\) has energy norm bounded by \(C/(1+|\operatorname{Im}p|)\), as shown in Appendix~\ref{app:radial-preparation}. The transform of the finite pulse is square integrable, so the contour integral obeys \eqref{eq:full-remainder}. By the Cauchy--Schwarz inequality for the energy norm, both the remainder energy and its cross term with the leading field decay faster than the cluster energy and are therefore subleading.

Set \(s=\gamma_*t\). Equations~\eqref{eq:cluster-energy} and \eqref{eq:full-remainder} give
\begin{equation}
\begin{alignedat}{2}
D_S&=a_Se^{-s}[1+o(1)],\quad & a_S&=\sqrt{\frac{\gamma_*}{2E_u}}|A_{H,S}|,\\
D_R&=a_Rs e^{-s}[1+O(s^{-1})],\quad & a_R&=\frac{|B_{H,R}|}{\sqrt{2\gamma_*E_u}}.
\end{alignedat}
\label{eq:late-amplitudes}
\end{equation}
The energy ratio \(\mathcal E_R/\mathcal E_S\sim(a_R/a_S)^2s^2\) therefore grows without bound. The initial order reverses and remains reversed at late times. Monotone energy loss then implies that \(S\) reaches every sufficiently small common threshold earlier and remains below it.

Writing \(\Lambda_\theta=\log(1/\theta)\) and taking logarithms of \eqref{eq:late-amplitudes}, we iterate the threshold condition for \(\tau_R\) to obtain
\begin{equation}
\begin{aligned}
\gamma_*\tau_S&=\Lambda_\theta+\log a_S+o(1),\\
\gamma_*\tau_R&=\Lambda_\theta+\log\Lambda_\theta+\log a_R
+O\!\left(\frac{\log\Lambda_\theta}{\Lambda_\theta}\right).
\end{aligned}
\label{eq:individual-clocks}
\end{equation}
The two responses reach the threshold at times set by the same leading scale \(\Lambda_\theta/\gamma_*\). The extra factor of time in the reference response must also be overcome by exponential decay, producing the additional \(\log\Lambda_\theta\). Corrections from the faster radial modes vanish as the threshold tends to zero.

The source moments determine the ratio of these amplitudes. The filtered source has \(A_{H,S}=-\mathsf b_*M_1[J_S]/g\), while \(B_{H,R}=\mathsf b_*M_0[J_R]/g\). Substituting these expressions into \(a_R/a_S=|B_{H,R}/(\gamma_*A_{H,S})|\) gives
\begin{equation}
\begin{gathered}
\gamma_*(\tau_R-\tau_S)=\log\Lambda_\theta+C_{SR}
+O\!\left(\frac{\log\Lambda_\theta}{\Lambda_\theta}\right),\\
C_{SR}=\log\left|\frac{M_0[J_R]}{\gamma_*M_1[J_S]}\right|.
\end{gathered}
\label{eq:moment-clock-offset}
\end{equation}
The filtered leading field and the term proportional to \(t\) in the reference response share the same coalesced radial mode. The common energy quadratic form weights this direction equally in both preparations, so its normalization, the pole coefficient and \(E_u\) cancel from the amplitude ratio. The double logarithm comes from the change in polynomial degree. The same dependence on the degree appears in the cutoff time for stochastic thermalization with weak noise \cite{BarreraHoegelePardo2021Cutoff,BarreraJara2020Thermalisation}. Here the weighted moments of the two boundary pulses give the finite offset before the recovery curves are computed.

Both recovery times grow as \(\Lambda_\theta/\gamma_*\), while their difference grows only as \(\log\Lambda_\theta/\gamma_*\). Their relative difference therefore obeys \((\tau_R-\tau_S)/\tau_R\sim\log\Lambda_\theta/\Lambda_\theta\to0\), with \(\tau_R/\tau_S\to1\). Including the common preparation duration leaves the absolute time saved unchanged and reduces its fraction of the total time at finite thresholds.

Before the asymptotic regime is reached, the exact cluster energy gives a finite time prediction that retains both contributions to the reference horizon field. Let \(\tau_{{\rm cl},i}\) be its threshold time and define \(\Delta\tau_{\rm cl}=\tau_{{\rm cl},R}-\tau_{{\rm cl},S}\) and \(\zeta_R=\gamma_*A_{H,R}/B_{H,R}-1/2\). For \(0<\theta<\min_{i=S,R}\sqrt{\mathcal E_{{\rm cl},i}(0)/E_u}\), both thresholds are reached at positive times. Completing the square in \eqref{eq:cluster-energy} then gives
\begin{equation}
\gamma_*\Delta\tau_{\rm cl}=C_{SR}+
\frac12\log\left[(\gamma_*\tau_{{\rm cl},R}+\zeta_R)^2+\frac14\right].
\label{eq:finite-cluster-clock}
\end{equation}
For the complete field, this gives the leading prediction once the faster remainder has decayed. The simpler asymptotic law requires \(\Lambda_\theta\) to be large compared with \(|\zeta_R|\) and the logarithms of the fixed amplitudes. A small but nonzero reference memory can delay that regime, making the finite time relation useful before the asymptotic offset is visible.

\subsection{Energy reversal for matched pulses}
\label{sec:fixed-example}
We test the finite time prediction on the fixed background used to construct the matched pulses. It lies close to the EP, as checked independently in Appendix~\ref{app:source-construction}. The source moments, the constraint \(M_0[J_S]=0\) and the EP cluster prediction use the reference rate \(\gamma_{\rm ref}\) in place of \(\gamma_*\). The complete field evolves on the actual background and retains its small spectral splitting. Both pulses have duration \(\tau_p=q^{-1}\), with \(\Delta=\ell_b=0.1q^{-1}\) and \(A_0=\epsilon q\). The linear recovery amplitude \(D_J\) is invariant under a common nonzero rescaling of the pulses.

\begin{figure*}[t]
\centering
\begin{minipage}[t]{.370\textwidth}\vspace{0pt}
{\setlength{\panelcaptionleft}{0.225963\linewidth}
\setlength{\subfigcapskip}{\dimexpr 3pt-0.031916\linewidth\relax}
\subfigure[Applied source]{\includegraphics[width=\linewidth]{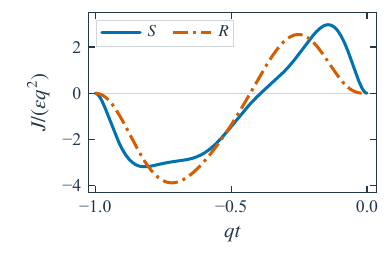}\label{fig:F2a}}}
\par\vspace{1.5pt}
{\setlength{\panelcaptionleft}{0.225963\linewidth}
\setlength{\subfigcapskip}{\dimexpr 3pt-0.031916\linewidth\relax}
\subfigure[Source moment]{\includegraphics[width=\linewidth]{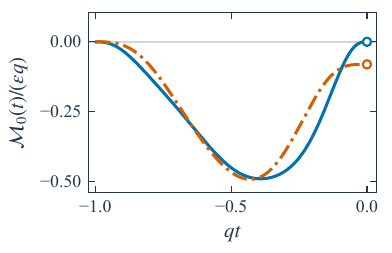}\label{fig:F2b}}}
\end{minipage}
\hfill
\begin{minipage}[t]{.610\textwidth}\vspace{0pt}
{\setlength{\panelcaptionleft}{0.116152\linewidth}
\setlength{\subfigcapskip}{\dimexpr 3pt-0.019359\linewidth\relax}
\subfigure[Energy recovery]{\includegraphics[width=\linewidth]{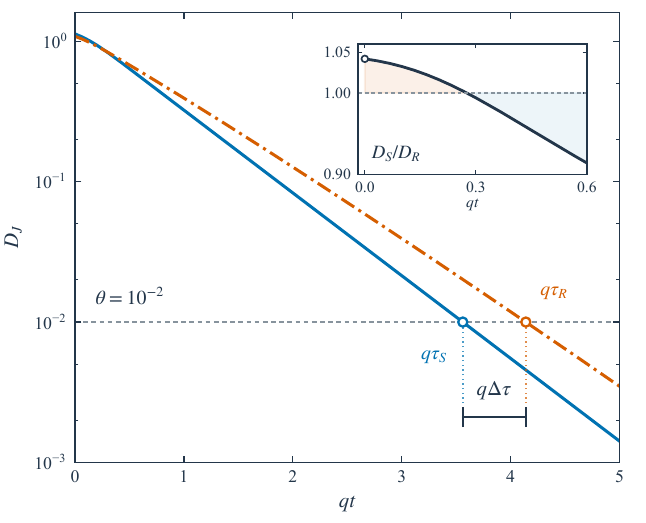}\label{fig:F2c}}}
\end{minipage}
\caption{Energy reversal for a fixed Robin pair with common preparation duration, task, source cost and injected work. Panel~\subref{fig:F2a} shows the applied pulses. Panel~\subref{fig:F2b} shows the accumulated moments \(\mathcal M_0(t)=\int_{-\tau_p}^{t}e^{\gamma_{\rm ref}s}J(s)ds\). The moment of \(S\) returns to zero at release. Panel~\subref{fig:F2c} shows the complete energy amplitude \(D_J=\sqrt{\mathcal E/E_u}\). Preparation \(S\) retains more energy at release but reaches \(\theta=10^{-2}\) earlier by \(q\Delta\tau=q(\tau_R-\tau_S)\). The inset resolves the reversal of \(D_S/D_R\). The legend in~\subref{fig:F2a} applies throughout.}
\label{fig:fair}
\end{figure*}

The pulses use a common basis of six functions, with both the source and its first derivative vanishing at release. Source \(S\) minimizes the cost under the task and moment constraints within this family. We choose \(R\) with the same task, cost and work but a nonzero zeroth moment. Both waveforms remain fixed throughout the threshold scan.

Table~\ref{tab:preparation-balance} and Fig.~\ref{fig:fair} show how this matched pair realizes the energy reversal near the EP, illustrating the mechanism analyzed at exact coalescence in Sec.~\ref{sec:recovery-times}. Source \(S\) loses less of the common work to the horizon during preparation and therefore leaves more energy at release. It also brings its zeroth source moment close to zero while retaining \(M_1[J_S]\ne0\). The complete energy curves subsequently reverse their order, and \(S\) reaches the displayed threshold sooner. The reversal also persists in a neighborhood of this example. With the pulses calibrated on each nearby fixed background, the continuous energy bounds in Appendix~\ref{app:source-construction} give \(q(\tau_R-\tau_S)>0.25\) at \(\theta=10^{-2}\) for relative source errors up to \(10^{-4}\) that preserve the common task, cost and work.

\begin{table}[htbp]
\centering
\begin{ruledtabular}
\begin{tabular}{lrr}
Quantity & \(S\) & \(R\)\\
\hline
\(Q/(\epsilon q)\) & \(1\) & \(1\)\\
\(\mathcal B/(\epsilon^2q^3)\) & \(6.94435\) & \(6.94435\)\\
\(\mathcal W/(\epsilon^2q^3)\) & \(5.53026\) & \(5.53026\)\\
\hline
\(M_0/(\epsilon q)\) & \(\simeq0\) & \(-0.0808085\)\\
\(M_1/\epsilon\) & \(0.259594\) & \(0.237415\)\\
\(\mathcal E(0)/(\epsilon^2q^3)\) & \(0.222832\) & \(0.205310\)\\
Preparation absorption & \(5.30743\) & \(5.32495\)\\
\end{tabular}
\end{ruledtabular}
\caption{Preparation balance for the fixed sources.
Preparation absorption is the horizon flux integrated from
\(-\tau_p\) to zero, in units of \(\epsilon^2q^3\).
The source moments are evaluated at \(\gamma_{\rm ref}\).
Entries are rounded.}
\label{tab:preparation-balance}
\end{table}

\begin{figure*}[t]
\centering
\begin{minipage}[t]{.328\textwidth}\vspace{0pt}
{\setlength{\panelcaptionleft}{0.235000\linewidth}
\setlength{\subfigcapskip}{\dimexpr 3pt-0.035613\linewidth\relax}
\subfigure[Time saved]{\includegraphics[width=\linewidth]{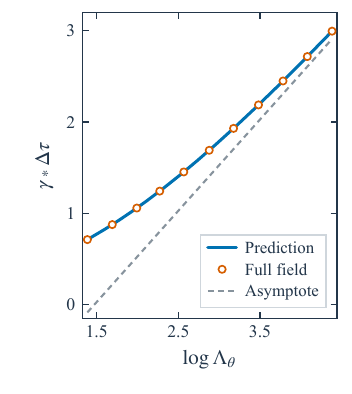}\label{fig:F3a}}}
\end{minipage}
\hfill
\begin{minipage}[t]{.328\textwidth}\vspace{0pt}
{\setlength{\panelcaptionleft}{0.235000\linewidth}
\setlength{\subfigcapskip}{\dimexpr 3pt-0.035613\linewidth\relax}
\subfigure[Moment offset]{\includegraphics[width=\linewidth]{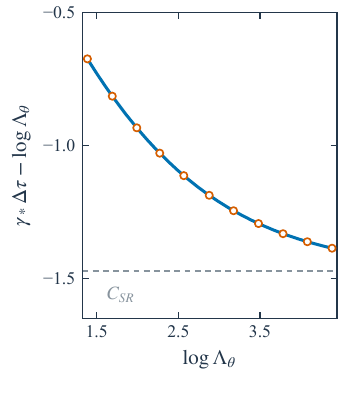}\label{fig:F3b}}}
\end{minipage}
\hfill
\begin{minipage}[t]{.328\textwidth}\vspace{0pt}
{\setlength{\panelcaptionleft}{0.235000\linewidth}
\setlength{\subfigcapskip}{\dimexpr 3pt-0.035613\linewidth\relax}
\subfigure[Relative gain]{\includegraphics[width=\linewidth]{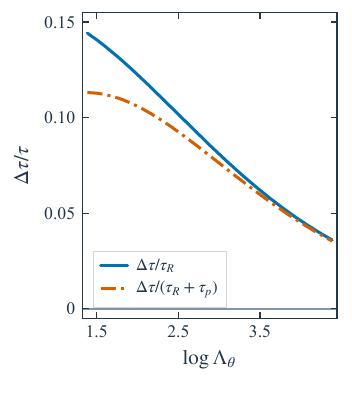}\label{fig:F3c}}}
\end{minipage}
\caption{Recovery times for the fixed waveforms in Fig.~\ref{fig:fair}. Panel~\subref{fig:F3a} compares the time saved with the finite cluster prediction \eqref{eq:finite-cluster-clock} and its logarithmic asymptote. The same legend applies to~\subref{fig:F3b}, where subtracting \(\log\Lambda_\theta\) reveals the approach to \(C_{SR}\), which is fixed by the source moments. Panel~\subref{fig:F3c} shows the relative gain after release, \(\Delta\tau/\tau_R\), and with preparation included, \(\Delta\tau/(\tau_R+\tau_p)\). The numerical panels evaluate \(\gamma_*\) at \(\gamma_{\rm ref}\). The comparison is restricted to the interval in which the small spectral splitting remains unresolved.}
\label{fig:clocks}
\end{figure*}

Figure~\ref{fig:clocks}\subref{fig:F3a} compares the time gained with the finite time prediction in \eqref{eq:finite-cluster-clock}. This prediction follows the complete field before the logarithmic asymptote becomes accurate. Panel~\subref{fig:F3b} subtracts \(\log\Lambda_\theta\) to reveal the approach to \(C_{SR}\simeq-1.47210\), as determined from the source moments. Panel~\subref{fig:F3c} displays the decreasing relative gain derived above, including the reduction when preparation time is counted.

All displayed numerical energies include the complete radial field and the Robin boundary term at \(N=48\). The marked crossings in Fig.~\ref{fig:fair} are obtained by interpolation in the sampled \(\log D_J\) trajectories. The cluster prediction uses source moments and independently calculated radial response coefficients. Calculations at \(N=32,48,64\) agree over the sampled recovery interval. Appendix~\ref{app:source-construction} specifies the background, source coefficients, evolution method and numerical checks. The scan extends to large \(\Lambda_\theta\) to test the approach to the asymptotic law in \eqref{eq:moment-clock-offset} on a fixed background. The small pole splitting remains unresolved throughout the scan. When the work of preparation also heats the black hole, the range of thresholds that can exhibit the EP law is limited by the conditions in Sec.~\ref{sec:detuning}.

\Needspace{7\baselineskip}
\section{Detuning and thermal feedback}
\label{sec:detuning}
\begingroup\widowpenalty=10000
We first examine how the linear recovery law changes on a nearby stationary background. The leading gravitational response then relates the work of preparation to the shift of the final spectrum and the remaining entropy production.
\par\endgroup

\subsection{Accumulated memory and resolved splitting}
In open quantum systems, imperfect state preparation can restore suppressed slow modes and reduce the relaxation advantage \cite{MackinnonPaternostro2026Robustness}. Here, preparation error can restore a time term, while background drift can separate the merged poles. We consider nearby preparations that retain their common task, cost and work, together with the strict initial energy order. Their actual coefficients \(A_{H,S}\) and \(B_{H,R}\) remain close to the fixed nonzero values of the EP pair. We also require a uniform bound on the faster radial remainder, with a common decay rate above the leading pair.

At coalescence, a residual \(B_{H,S}\) changes the filtered horizon field to
\(e^{-\gamma_*t}(A_{H,S}+B_{H,S}t)\). Relative to the ordinary exponential, the restored term grows as \(\eta_M\gamma_*t\), where
\(\eta_M=|B_{H,S}|/(\gamma_*|A_{H,S}|)\).
A pure gain error preserves an exact zero of \(M_0\) but changes the task, cost and work, whereas a shape error can restore the missing moment. Thus \(\eta_M\) measures the residual time term rather than an arbitrary waveform error.

Varying the stationary Robin background separates the leading poles. Holographic pseudospectra describe their sensitivity more generally \cite{AreanEtAl2023Pseudospectra}. Near the merger, individual excitation coefficients can grow while the paired response remains regular \cite{NakamotoOshita2026Nariai}.

For the two leading poles \(p_\pm=-\gamma\pm d\), we define the dimensionless detuning \(\xi=d^2/\gamma_*^2\). The corresponding horizon field is
\begin{equation}
y(t)=e^{-\gamma t}\left[A_H\cosh(dt)+B_H\frac{\sinh(dt)}{d}\right].
\label{eq:split-field}
\end{equation}
As \(d\to0\), the field reduces smoothly to \(e^{-\gamma t}(A_H+B_Ht)\). The center rate satisfies \(\gamma=\gamma_*+O(\xi)\). Positive \(\xi\) separates the decay rates on the real axis. Negative \(\xi\) gives a complex conjugate pair, with frequency \(\Omega=\gamma_*\sqrt{-\xi}\). For \(|d|t\ll1\), expansion of the hyperbolic functions changes the polynomial envelope by order \(|d|^2t^2\). The two modes become resolved when \(|d|t\sim1\). Away from coalescence, \(\eta_M\) is defined from the actual coefficients in \eqref{eq:split-field}.

The equation of motion for the mode pair determines its remaining energy through the future horizon flux,
\begin{equation}
\begin{aligned}
\ddot y+2\gamma\dot y+(\gamma^2-d^2)y&=0,\\
\mathcal E_{\rm cl}&=\frac{\dot y^2+(\gamma^2-d^2)y^2}{4\gamma}.
\end{aligned}
\label{eq:split-energy}
\end{equation}
This energy is positive for the stable pair near the EP on either side of the splitting. Its derivative is \(-\dot y^2\), and it remains regular when the horizon field vanishes.

Recovery occurs at \(\gamma_*t\sim\Lambda_\theta\). The accumulated memory \(\eta_M\gamma_*t\) and splitting correction \(|d|^2t^2\) therefore remain small when
\begin{equation}
\eta_M\Lambda_\theta\ll1,\qquad
|\xi|\Lambda_\theta^2\ll1.
\label{eq:protected-window}
\end{equation}
Once the faster modes have decayed, these conditions keep the energy close to the EP prediction. At late times the energy falls at a finite fractional rate, so a small relative change in energy produces only a small shift in the time at which it reaches the threshold. Equation~\eqref{eq:protected-window} gives a sufficient observation window for the fixed pair. Maintaining this window at a deeper threshold calls for a smaller residual moment and smaller background detuning.

At exact coalescence, a fixed nonzero residual \(M_0[J_S]\) eventually restores the linear time factor in the filtered response. For fixed pulses with \(M_0[J_R]\ne0\), the late time inversion gives
\[
\lim_{\theta\to0}\gamma_*(\tau_R-\tau_S)
=\log\left|\frac{M_0[J_R]}{M_0[J_S]}\right|.
\]
The growing advantage becomes a constant offset, which remains positive when \(|M_0[J_S]|<|M_0[J_R]|\). For a detuned background, the resolved modes in \eqref{eq:split-field} instead determine the late recovery.

\subsection{Work and the final spectrum}
We now allow the geometry to respond to the scalar. We take the bulk action to be
\[
S_{\rm bulk}=\int d^4x\sqrt{-g}\left[\frac{R+6/L^2}{16\pi G}-\frac{Z_\Phi}{2}\bigl((\nabla\Phi)^2+\mu^2\Phi^2\bigr)\right],
\]
with \(\mu^2=-2/L^2\) and \(Z_\Phi>0\). We keep the same Robin boundary condition and a fixed boundary metric. In the boundary expansion \(\Phi=z\alpha+z^2\beta+\cdots\), the scalar kinetic term fixes the Robin work normalization to \(\mathcal N=Z_\Phi L^2\).

A pulse of amplitude \(\epsilon\) sources a scalar field at order \(\epsilon\). In the absence of an independent gravitational perturbation, the metric responds first at order \(\epsilon^2\), because the scalar stress is quadratic. The pulse and the initial black hole preserve translations and rotations in the transverse plane, so there is no independent tensor mode. In ingoing EF coordinates with the transverse area used to fix the radial coordinate, the radial Einstein constraints determine the metric on each null slice given the scalar profile and the boundary energy density. The Ward identity fixes the change in that density from the work of the source. We assume that the solution remains regular and settles to a planar Schwarzschild AdS black hole with no scalar hair after release.

As in holographic quenches \cite{BuchelEtAl2013Quenches,ChenEtAl2023ThermalQuench}, the boundary work fixes the energy of the final black hole. For the stationary family, the first law \(d\mathfrak m=T\,ds\) with \(s=L^2/(4Gz_h^2)\) gives the mass density. With the zero of energy at \(T=0\), the density and endpoint balance are
\begin{equation}
\mathfrak m(T)=\frac{8\pi^2L^2}{27G}T^3,\qquad
\mathfrak m(T_f)-\mathfrak m(T_i)=w,
\label{eq:endpoint-balance}
\end{equation}
where \(w\) is the full injected work per transverse area. A common initial black hole and equal work give the same final temperature. The preparations in Table~\ref{tab:preparation-balance} have equal work at quadratic order to numerical precision, and hence the same mass shift at that order. Matching the full work at finite amplitude would require accounting for the effect of the evolving geometry during preparation. At leading order, the work balance \eqref{eq:work-partition} and the task bound established after \eqref{eq:recovery-amplitude} imply \(\mathcal W\ge\mathcal E_0\ge E_u>0\). Thus a nonzero prescribed response requires positive work and heats the black hole.

The temperature changes the dimensionless Robin coupling \(\bar\kappa=\kappa_{\rm phys}z_h=3\kappa_{\rm phys}/(4\pi T)\) used in Ref.~\cite{R1}. At fixed physical coupling, radial scaling gives \(m(p,T)=\lambda m(p/\lambda,T_*)\), where \(\lambda=T/T_*\). Expanding the pole equation about \(m=q\) and \(m'=0\) gives
\[
\begin{gathered}
\xi=-\Gamma_{\rm EP}\frac{T-T_*}{T_*}
+O\!\left(\frac{(T-T_*)^2}{T_*^2}\right),\\
\Gamma_{\rm EP}=\frac{2q}{\gamma_*^2m''(-\gamma_*)}>0.
\end{gathered}
\]
A positive temperature shift therefore moves the final spectrum to the side with a complex conjugate pair.

Let \(c_*=(d\mathfrak m/dT)_{T_*}\) denote the heat capacity per transverse area at the EP temperature. Suppose preparation begins at \(T_i=T_*\), with weak heating \(w/(c_*T_*)\ll1\). The endpoint balance and the detuning relation give
\begin{equation}
\begin{aligned}
\frac{T_f-T_*}{T_*}
&=\frac{w}{c_*T_*}
+O\!\left[\left(\frac{w}{c_*T_*}\right)^2\right],\\
\frac{\Omega_f^2}{\gamma_*^2}
&=\Gamma_{\rm EP}\frac{w}{c_*T_*}
+O\!\left[\left(\frac{w}{c_*T_*}\right)^2\right].
\end{aligned}
\label{eq:heating-splitting}
\end{equation}
The actual injected work fixes the distance of the final spectrum from coalescence. For a fixed pulse shape, \(w=\epsilon^2\mathcal N\mathcal W_0+O(\epsilon^4)\), with \(\mathcal W_0>0\). The temperature rise is quadratic in source amplitude, and the square root structure of the EP gives \(\Omega_f\propto|\epsilon|\) at leading order. The center rate changes only quadratically. 

For linear response about the final state, a sufficient condition for unresolved splitting is
\[
\frac{\Gamma_{\rm EP}w}{c_*T_*}\,\Lambda_\theta^2\ll1.
\]
For a fixed pulse shape, substituting the leading quadratic work gives
\[
\Lambda_\theta\ll\frac{1}{|\epsilon|}
\sqrt{\frac{c_*T_*}{\Gamma_{\rm EP}\mathcal N\mathcal W_0}}.
\]
Reducing \(|\epsilon|\) in the example also reduces the prescribed response \(A_0=\epsilon q\). For a prescribed leading response \(A_0\), the quadratic work density is at least \(\mathcal N K_{\rm stat}A_0^2/2\), where \(K_{\rm stat}\) is evaluated at \(T_*\). At leading order, keeping the final splitting unresolved at recovery then requires
\[
|A_0|\Lambda_\theta \ll
\sqrt{\frac{2c_*T_*}{\Gamma_{\rm EP}\mathcal N K_{\rm stat}}}.
\]
Thus a fixed nonzero task limits how far pulse shaping alone can extend the EP recovery law. The actual work can impose a tighter limit, and the residual source moment still has to satisfy \eqref{eq:protected-window}.

The final spectrum can be brought to the EP by choosing a cooler initial black hole with
\(\mathfrak m(T_i)=\mathfrak m(T_*)-w\).
For positive work, strict initial stability requires \(w<\mathfrak m(T_*)-\mathfrak m(T_c)\).
For this choice, an error in the actual work adds to the initial mass error. A sufficient condition for keeping the final splitting unresolved is
\[
\Gamma_{\rm EP}\frac{|\delta\mathfrak m_i+\delta w|}{c_*T_*}
\Lambda_\theta^2\ll1.
\]
In this way, matching the final mass fixes the stationary spectrum, whereas tuning the source moment controls the polynomial coefficient of the released linear field.

\subsection{Remaining energy and entropy production}
At the first nonzero gravitational order, horizon absorption also governs how entropy production is distributed between preparation and free relaxation. We write \(f_J=\epsilon f_J^{(1)}+O(\epsilon^3)\) and use the superscript \((2)\) for the coefficient of \(\epsilon^2\). The event horizon generator \(k^\mu\) is normalized by boundary time and has background surface gravity \(\varkappa_H=2\pi T\). Planar isotropy makes the shear vanish. The expansion starts at order \(\epsilon^2\), so its square contributes only at order \(\epsilon^4\).

The scalar contributes \(T_{kk}=Z_\Phi(k^\mu\partial_\mu\Phi)^2\) to null focusing. On the background horizon, \(\Phi_H^{(1)}=z_h f_H^{(1)}\), so the quadratic stress weighted by the area density \(L^2/z_h^2\) is \(\mathcal N|\dot f_H^{(1)}|^2\). Inserting this flux in the Raychaudhuri equation and dividing the area density by \(4G\) gives the focusing equation for the event horizon entropy density,
\[
\ddot s_{\rm EH}^{(2)}-\varkappa_H\dot s_{\rm EH}^{(2)}
=-2\pi\mathcal N|\dot f_H^{(1)}|^2.
\]
The local dynamical entropy of Refs.~\cite{H01,H02} then obeys \(T\dot s_{\rm dyn}^{(2)}=\mathcal N|\dot f_H^{(1)}|^2\). Integrating to the stationary endpoint and using \eqref{eq:remaining-absorption} after release gives
\begin{equation}
\begin{gathered}
s_{\rm dyn}^{(2)}=s_{\rm EH}^{(2)}-\varkappa_H^{-1}\dot s_{\rm EH}^{(2)},\\
T[s_\infty^{(2)}-s_{\rm dyn}^{(2)}(t)]=\mathcal N\mathcal E[f_J^{(1)}(t)].
\end{gathered}
\label{eq:entropy-deficit}
\end{equation}
The dynamical entropy coefficient equals that of the apparent horizon on the associated slices \cite{VisserYan2024DynamicalEntropy,FurugoriEtAl2025ApparentEntropy}. A common initial black hole and equal quadratic work give a common final entropy coefficient. At this order, the pulses redistribute the same total entropy production between preparation and recovery. For the pair exhibiting energy reversal, the filtered pulse loses less energy to the horizon during preparation. It leaves more exterior energy and a larger local dynamical entropy deficit at release, yet reaches a common remaining deficit sooner.

\section{Summary and outlook}
\label{sec:discussion}
In this work, we have shown how finite boundary preparation can change the order of recovery near a black hole EP without altering the leading decay rate. At the exceptional point, the linear time factor disappears from the leading response when the zeroth weighted source moment vanishes and the first remains nonzero, leaving an ordinary exponential. At late times this filtered response has a higher fractional absorption rate than a reference that retains the time factor. The difference decreases as \(2/t\), but its accumulated effect can allow a perturbation with more release energy to recover sooner. At the exact EP, the resulting recovery time advantage for a suitable fixed pair grows as \(\gamma_*^{-1}\log\log(1/\theta)\) as the common threshold \(\theta\) is lowered. Full scalar field evolution near the EP illustrates the reversal for pulses with matched preparation resources and equal injected work.

The source moments also determine the constant offset in the asymptotic time difference. The two leading responses share the same coalesced radial direction, so the radial normalization and pole coefficient cancel from their amplitude ratio. This result connects the preparation history to the late recovery time. As the threshold decreases, the absolute time advantage grows, while its fraction of the total recovery time tends to zero.

In the black hole model considered here, a boundary pulse both excites the perturbation and injects energy into the system. Even if two preparations inject the same work, the horizon can absorb different amounts during preparation, leaving different energies at release. When the leading gravitational backreaction is included, we find that the injected work also heats the black hole and changes its final temperature and mode spectrum. If preparation begins at the EP, the temperature rise caused by a weak drive is quadratic in the source amplitude, while the final spectral splitting can appear at linear order. This splitting limits the observation window for the EP recovery law. To observe this exceptional point scaling, the faster modes must have decayed while the effect of the splitting remains small.

An experimental test could begin by tuning one pulse until the linear time factor disappears from its leading response. With that pulse and a reference pulse held fixed, one could then measure how the time saved varies with the recovery threshold. Controlled dissipative systems such as coupled pendulums, superconducting resonators, and qubits offer possible platforms for studying related responses \cite{EPpendulum,EPresonators,LejeuneEtAl2026Reset}. When changes to the dissipative background caused by a drive are negligible, such platforms may allow the fixed background scaling to be tested separately from feedback effects. In the gravitational setting, by contrast, preparation and recovery at finite amplitude require the coupled evolution of the background and scalar field. It would be valuable to examine whether the initial black hole temperature and the boundary pulse can be chosen together to control both the final spectrum and mode excitation under a prescribed task and resource constraints. This would clarify how the recovery advantage can be realized in the nonlinear regime.

\begin{samepage}
\begin{acknowledgments}
This work was supported by the National Natural Science Foundation of China under Grants No. 12305063 and No. 12547143.
\end{acknowledgments}
\end{samepage}

\appendix
\section{Radial response and fixed boundary preparation}
\label{app:radial-preparation}
\subsection{The stable leading double pole}
We give the radial argument underlying Sec.~\ref{sec:source-memory}. The response function and the horizon normalization are those of \eqref{eq:response}.

The poles of \eqref{eq:response} give the quasinormal frequencies in the Laplace convention. The exponential horizon tail \(V\sim e^{-\nu x}\), with \(\nu=3/z_h\), permits analytic continuation to \(\operatorname{Re}p>-\nu/2\) \cite{R12}. The radial solution obeys the integral equation
\[
\begin{aligned}
G_p(x)&=1+\int_x^\infty K_p(y-x)V(y)G_p(y)\,dy,\\
K_p(s)&=\int_0^s e^{-2pr}\,dr .
\end{aligned}
\]

The positive potential explains why the merger governs the late response. For real \(p=-a\) in this strip, the radial Wronskian gives
\begin{equation}
\begin{gathered}
S(a)=\int_0^\infty e^{2ax}\bigl(G_{-a}^2-1\bigr)dx,\\
m'(-a)=\frac{1-2aS(a)}{G_{-a}(0)^2}.
\end{gathered}
\label{eq:radial-identity}
\end{equation}
Iterating the integral equation with \(V>0\) and \(K_{-a}>0\) gives \(G_{-a}>1\) and \(\partial_a G_{-a}>0\), which yields \(S>0\) and \(S'>0\). The nonzero horizon tail makes \(S\) diverge as \(a\uparrow\nu/2\). There is consequently one solution \(a=\gamma_*\) of \(2aS(a)=1\). At this point
\(m''(-\gamma_*)=[2S+2\gamma_*S']/G_{-\gamma_*}(0)^2>0\).
Setting \(\kappa_*=-m(-\gamma_*)\) produces a visible double pole.

Having established the double pole on the negative real axis, we examine whether a slower mode could dominate the late response. For a nonreal decaying resonance \(p=-a+i\omega\), the continued flux identity gives
\(\int_0^\infty e^{2ax}(|G_p|^2-1)dx=1/(2a)\).
Strict comparison in the same integral equation yields
\(|G_{-a+i\omega}|<G_{-a}\). Therefore \(2aS(a)>1\), so \(a>\gamma_*\). On the real axis, \eqref{eq:radial-identity} gives \(m(p)>m(-\gamma_*)\) for \(-\gamma_*<p\le0\), excluding a slower real pole. In particular, \(\kappa_*>\kappa_c=-m(0)\). Positive energy excludes growth, and the boundary Wronskian
\(\operatorname{Im}[\overline{G_{i\omega}(0)}H_\kappa(i\omega)]=-\omega\)
excludes a nonzero imaginary pole. The double pole is therefore stable and leading.

To determine the sign of the Robin coupling at the merger, we return to the positive integral equation. In agreement with Ref.~\cite{R1}, its first iterate in units \(z_h=1\) gives
\[
S(1/2)\ge1+2\int_0^1z[x(z)-1]e^{x(z)}dz>1.
\]
The last inequality follows from \(x(z)\ge-\log(1-z)/3+4z/9\). Since \(2aS(a)\) increases, its unit crossing obeys \(\gamma_*<1/2\). Writing \(I_0=\int VG_{-a}dx\) and \(I_2=\int e^{2ax}VG_{-a}dx\), the boundary equation gives \(m(-a)G_{-a}(0)=(I_2+I_0)/2-a\), with \(I_2\ge I_0>1/2\). Hence \(m(-\gamma_*)>0\) and \(\kappa_c<\kappa_*<0\). At fixed negative physical coupling the merger temperature \(T_*\) lies above \(T_c=3q/(4\pi|\bar\kappa_c|)\). The double pole has a finite decay rate, while the zero mode occurs at \(T_c\). Memory at this merger can therefore be studied within the stable phase.

The same radial integral equation gives the estimate in \eqref{eq:full-remainder}. On a line \(p=-a_r+i\omega\) in the spectral gap, Volterra comparison bounds \(|G_p(x)|\) uniformly and gives \(|G_p'(x)|\le C e^{-\nu x}\). At large \(|\omega|\), the explicit kernel gives \(G_p(0)=1+O(|\omega|^{-1})\), so \(H_\kappa(p)=-p+O(1)\). On bounded frequency intervals the denominator stays away from zero. Together with \eqref{eq:energy}, these estimates give
\[
\left\|G_p/H_\kappa(p)\right\|_{\mathcal E}
\le \frac{C}{1+|\omega|}.
\]
The Volterra bounds hold uniformly on closed substrips and give analyticity of \(G_p\) in the energy norm. For smooth pulses of compact support, we shift the inverse Laplace contour across the double pole to this line. The source transform there is the Fourier transform of \(e^{a_rs}J(s)\). By Plancherel and the Cauchy--Schwarz inequality, the remaining integral is bounded in energy norm by \(C\|J\|_{L^2}e^{-a_rt}\). Density extends the estimate to square integrable sources supported in the preparation interval.

\subsection{Source construction and complete field evolution}
\label{app:source-construction}
The fixed pair in Sec.~\ref{sec:fixed-example} is prepared on the background below. The second quantity is the rate used to define its source moments,
\begin{equation}
\begin{gathered}
qz_h=0.2868269680251661,\\
\gamma_{\rm ref}/q=1.3567083943014633.
\end{gathered}
\label{eq:example-background}
\end{equation}
We retain the binary64 background used to construct the fixed pulses and match their resources. An independent calculation using 200 terms of the horizon series at 55 decimal digits locates the continuum double root close to this background. Near a double root, the pole separation varies as the square root of the parameter displacement. The discretization can also split the poles. Our evolution uses the collocation generator as computed and retains its pole separation.

The source is supported on \([-\tau_p,0]\). We use
\begin{equation}
\begin{gathered}
J_i(t)=\epsilon q^2\sum_{n=0}^{5}c_{i,n}
\sin^2(\pi u_p)P_n(2u_p-1),\\
u_p=\frac{t+\tau_p}{\tau_p},\qquad i=S,R,
\end{gathered}
\label{eq:numerical-source-basis}
\end{equation}
where \(P_n\) is a Legendre polynomial. We set
\(\tau_p=q^{-1}\), \(\Delta=\ell_b=0.1q^{-1}\), and
\(A_0=\epsilon q\). Both the source and its first derivative vanish at
the endpoints. The amplitude \(\epsilon\) only fixes the overall scale
in the linear problem.

Within this source family, \(S\) minimizes the source cost subject to
\(Q=A_0\) and \(M_0=0\). The moment is evaluated at
\(\gamma_{\rm ref}\). We choose \(R\) with the same target, cost and
work, while allowing \(M_0\ne0\). Table~\ref{tab:source-coefficients}
lists the coefficients.

At radial resolution \(N=48\), both
targets agree with \(A_0\) to within \(1.4\times10^{-10}\) in the
units of Table~\ref{tab:preparation-balance}. The work difference is
below \(1.4\times10^{-10}\), and the cost difference is below
\(6\times10^{-14}\). These are equalities to numerical precision.

\begin{table}[htbp]
\centering
\small
\begin{ruledtabular}
\begin{tabular}{rrr}
\(n\) & \(c_{S,n}\) & \(c_{R,n}\)\\
\hline
0 & \(-0.4223815742388082\) & \(-2.0452862996366203\)\\
1 & \(34.79191109042825\) & \(10.63165990628891\)\\
2 & \(3.3309866071198506\) & \(-2.577044361030932\)\\
3 & \(30.989240255938352\) & \(-2.9961983353935864\)\\
4 & \(1.7119627447327916\) & \(-1.661312230104373\)\\
5 & \(10.083315189290373\) & \(-3.2064277991891377\)\\
\end{tabular}
\end{ruledtabular}
\caption{The fixed coefficients used in the numerical example.
The two sources share the basis in \eqref{eq:numerical-source-basis}.}
\label{tab:source-coefficients}
\end{table}

We discretize \(u=z/z_h\) by Chebyshev collocation, a standard method for black hole quasinormal mode calculations \cite{Jansen2017QNM}. The \(N+1\) nodes are
\(u_j=[1-\cos(\pi j/N)]/2\). The ingoing field equation is regular at
the horizon, and the boundary row imposes the driven Robin condition.
During preparation we integrate the resulting linear equations with the adaptive Runge--Kutta method DOP853, using relative and absolute tolerances of \(2\times10^{-12}\) and \(2\times10^{-14}\), respectively, in units with \(q=1\). We evaluate the energy from the radial interpolant by Gauss--Legendre quadrature, including the Robin boundary term.

For the threshold scan, let \(A_N\) denote the generator after release and let \(v_N\) be the numerical field at release. We factor out the common exponential and compute the compensated field
\[
u_N(t)=\exp\!\bigl[(A_N+\gamma_{\rm ref}I)t\bigr]v_N
\]
using a matrix exponential in binary64 arithmetic. The physical field is obtained by multiplying \(u_N\) by \(e^{-\gamma_{\rm ref}t}\). With \(E_N\) the discrete energy matrix, we use Brent's method to find the time at which the following expression equals \(\log\theta\).
\[
\log D_J(t)=\tfrac12\log[u_N(t)^{\mathsf T}E_Nu_N(t)/E_u]-\gamma_{\rm ref}t.
\]
This procedure carries the common decay in the logarithm and avoids repeated time steps of an exponentially small field after release.

For the cluster prediction we evaluate the EP formulas at
\(\gamma_{\rm ref}\). The radial calculation gives
\(\mathsf a_*\simeq-0.6017333057\) and
\(\mathsf b_*/q\simeq-4.3397584007\).
The boundary value and frequency derivative are
\(g\simeq1.7673765516\) and \(qh\simeq-0.3338598268\).
Together with the source coefficients, these fix \(A_H\) and \(B_H\)
through the map in Sec.~\ref{sec:source-memory}.
The common energy unit is \(E_u\simeq0.1753867000\) in units
\(\epsilon=q=1\).

We evolve the same sources at \(N=32,48,64\) and scan \(4\le\Lambda_\theta\le80\). At the sampled recovery times, the relative difference between the complete numerical energy and the cluster prediction is below \(1.1\times10^{-6}\). The comparison uses compensated energies, whose common exponential cancels from the ratio. The change in recovery time across resolutions is below \(3.6\times10^{-7}/q\). For the leading poles of each collocation generator, \(|p_+-p_-|\tau_{\max}<0.0025\), so the splitting remains unresolved over this interval.

For the robustness check, we take static backgrounds with \(qz_h\) within \(2^{-35}\) of the stored value in \eqref{eq:example-background}. The coefficients in Table~\ref{tab:source-coefficients} provide the starting point for calibration. On each background we minimize the cost of \(S\) under the task and moment constraints at the same \(\gamma_{\rm ref}\), then calibrate \(R\) to match the task, cost and work. The common cost, work and energy unit may vary across the interval. Around each calibrated pulse \(J_i\), we allow errors \(\delta J_i\in H_0^1(-\tau_p,0)\) with \(\sqrt{\mathcal B[\delta J_i]}\le10^{-4}\sqrt{\mathcal B[J_i]}\) that preserve those three constraints.

We bound the continuous field energy by correcting the degree 32 radial reconstructions to satisfy the Robin condition exactly and estimating the effect of their bulk residual with the positive energy identity. We use interval arithmetic to bound the time integral of the residual and the corrections from recalibrating the two pulses. For the execution errors, we use the work as the integrated horizon flux of the complete response. The real frequency Wronskian and \(|G_{i\omega}(0)|\le G_0(0)\) bound this work by \(G_0(0)^2\|\delta J_i\|_{L^2}^2\). The same bound applies to its exterior energy after release by \eqref{eq:work-partition} and monotonicity. We add these errors in the square root energy norm. Including all three bounds gives \(\mathcal E_S(0)>\mathcal E_R(0)\) and
\[
\begin{gathered}
qt_S=3.72802734375,\qquad q(t_R-t_S)=0.25,\\
D_S(t_S)<0.008905,\qquad D_R(t_R)>0.010888.
\end{gathered}
\]
Thus \(\tau_S\le t_S\) and \(\tau_R>t_R\) at the common threshold \(10^{-2}\). This gives a sufficient interval of background and source tolerances for the finite time reversal.

\end{document}